\documentclass[prb,aps,twocolumn]{revtex4}
\usepackage{amsmath}
\usepackage{amssymb}
\usepackage {graphicx}

\newcommand{\beq}{\begin{equation}}
\newcommand{\eeq}{\end{equation}}
\newcommand{\bea}{\begin{eqnarray}}
\newcommand{\eea}{\end{eqnarray}}

\begin{document}

\title{Manifesto for a higher Tc -- lessons from pnictides and
 cuprates}

\author{D.N. Basov$^1$ and  Andrey V. Chubukov$^2$}
\affiliation {$^1$ Department of Physics
9500 Gilman Drive, La Jolla, CA 92093-0319\\
$^2$ Department of Physics, University of Wisconsin,
1150 University ave., Madison, WI 53706}
\date{\today}
\begin{abstract}
We  explore energy scales, features in the normal state transport, relevant interactions and constraints for the pairing mechanisms in the high-Tc cuprates and Fe-pnictides. Based on this analysis we attempt to identify a
 number of attributes  
of  superconductors with a higher $T_c$.  Expanded version of the article published in Nature Physics, 7, 271 (2011).  
\end{abstract}
\maketitle


\section{Pnictides vs other exotic superconductors}

The discovery of superconductivity in $Fe$-based pnictides in 2008
(binary compounds of the elements from the 5th group: $N, P, As, Sb, Bi$)
  with $T_c$ reaching  almost
 $60K$  was, arguably, among the most significant breakthroughs in condensed matter physics during the past decade~\cite{discovery}.
  The excitement was enormous and so were the efforts --  the amount of data obtained for Fe-pnictides over the last three years  is
 comparable to that collected  for other known 
 superconductors over several decades~\cite{review,review_2,review_3}. 
 A large number of new superconducting materials have been discovered not 
only within the Fe-pnictide family but also in the Fe-chalkogenides group: Fe-based compounds with elements from the 16th group: $S, Se, Te$.   

Before 2008, the term ``high-temperature superconductivity'' (HTS) was
 reserved  for $Cu$-based superconductors (CuSC), discovered in 1986.  The 
 transformation from the ``Copper age'' to  the ``Iron age'' was swift and
 the term HTS  now equally applies to both CuSC
 and Fe-based superconductors (FeSC).

\begin{figure}[tbp]
\includegraphics[angle=0,width=\linewidth]{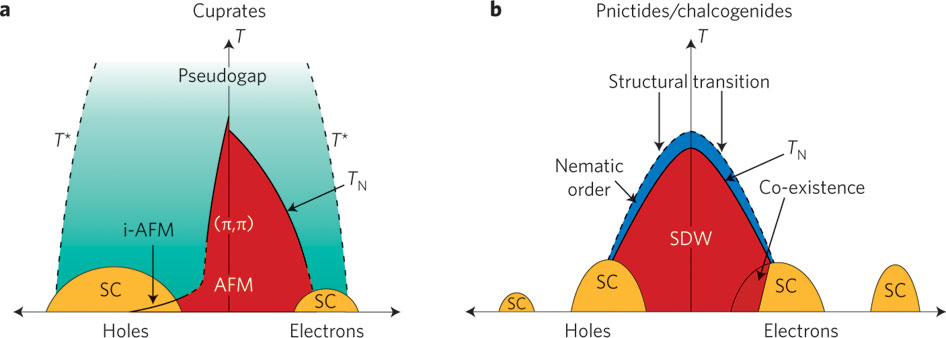}
\caption{Schematic phase diagrams of the cuprates and Fe-based pnictides upon hole or electron dopings~\cite{review,review_2,review_3}. In the shaded region, superconductivity and 
 antiferromagneism co-exist. Not all details/phases are shown.
Superconductivity in Fe-based systems can be initiated not only by doping but also by pressure and/or isovalent replacement of one pnictide element by another~\cite{nakai}.
 Nematic phase in pnictides at $T > T_N$ is subject of debates.
 Superconductors at large doping are  $KFe_2As_2$ for hole doping and $AFe_2Se_2$ ($A = K, Rb, Cs$) for electron doping. 
 Whether superconductivity in pnictides exists at all
 intermediate dopings is not clear yet.  
An additional superconducting dome in very strongly hole doped cuprates has also been reported~\protect\cite{gebale}}
\label{fig1}
\end{figure}
 Why  FeSC are such a big deal?  Even 
a cursory look at the phase diagram (Fig. \ref{fig1}) and the
properties of FeSC  reveal~\cite{review}
 an intricate interplay between magnetism and superconductivity, also typical for other ``exotic'' superconductors  discovered
  in the last three decades of the last millennium -- 
 heavy fermions, cuprates, ruthenates, 
 organic and molecular conductors. 
 Magnetism and superconductivity are
 antithetical in elemental superconductors,  but in exotic superconductors 
  magnetism associated with either d- or f-electrons is believed to be more a friend than a foe of the zero resistance state.
However, with the exception of the cuprates the $T_c$ of exotic superconductors  known before 2008  were quite low, and 
  many considered CuSC to being unique among exotic superconductors~\cite{lee_review}.
 The FeSC's with $T_c$'s comparable to some CuSC's,
  appear to undermine the uniqueness of the cuprates and have
 prompted the community to rethink what is important and what is not for the occurrence of high-$T_c$ superconductivity in any material. 
 Empowered by the {\it two} complementary perspectives on the high-Tc phenomenon one is well poised to address (and resolve) a number of outstanding issues
 such as:
(i) do all high-Tc materials  superconduct for the same reason? (ii)
 are the rather anomalous normal state properties of exotic superconductors 
 a necessary prerequisite for 
 high-Tc superconductivity? (iii) is there a generic route to increase $T_c$?
Below we give our perspective on these three issues.  
We argue that in CuSCs and FeSCc the answer to the first two questions is affirmative, and use the commonalities between the two classes of materials to detect
 the tools for the search for a higher $T_c$

We leave aside a number of interesting commonalities of CuSC and FeSC 
which only peripheraly related to superconductivity, including the origin of magnetic order and Fermi Surface (FS) reconstructruction in the magnetically ordered state, nematic order in FeSC above magnetic $T_N$ and its relation to 
 nematicity observed in the pseudogap phase in CuSC, and many others.
 A detailed review of the properties of Fe-based materials is given in Ref. \cite{review}, which also contains an extensive list of references.

\section{Phase diagrmas} 
 
\begin{figure}[tbp]
\includegraphics[angle=0,width=\linewidth]{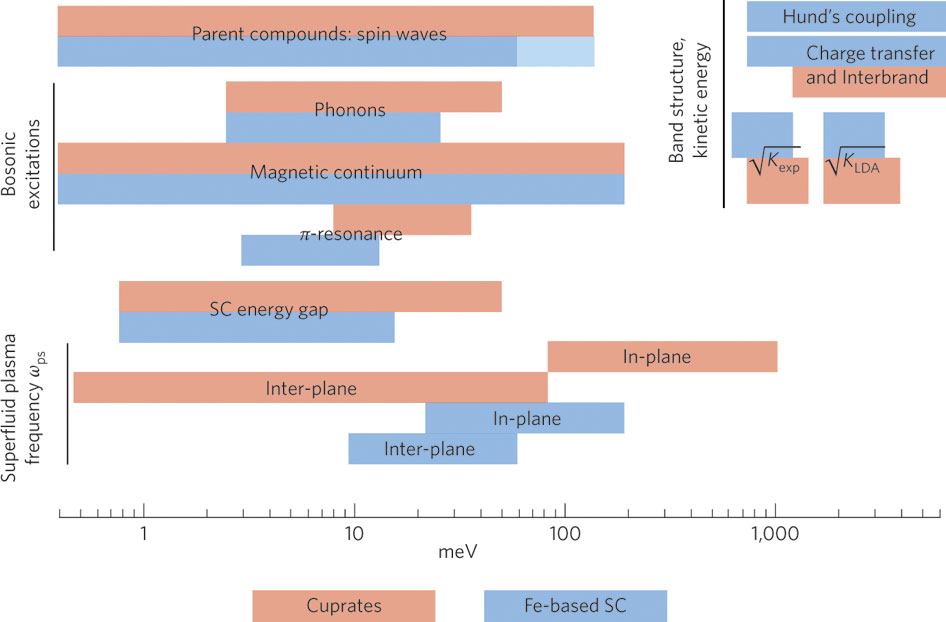}
\caption{Characteristic energy scales for fermionic and bosonic excitations, electronic kinetic energy $K_{exp}$, superfluid density  and superconducting energy gap in cuprates (red boxes) and Fe-based materials (blue boxes). $K_{LDA}$ is the band structure kinetic energy discussed in the text.}
\label{Table2}
\end{figure}

From a distance, phase diagrams and relevant energy scales
 of FeSC and CuSC look amaizingly similar (see Fig. \ref{fig1} 
and Table \ref{Table2}). 
In both classes of systems there is a region of a magnetic order near zero doping, and  superconductivity emerges upon either hole or electron doping.

On a closer look, however, there are notable 
differences. Magnetic order in undoped CuSC  is a conventional antiferromagnetism (spins of nearest neighbors are aligned antiparallel to each other), while  
 in most of undoped FeSC the order is antiferromagnetic in one direction and 
ferromagnetic in the other (a stripe order~\cite{review}). 
  The superconducting 
 order parameter in CuSC has $d-$wave symmetry, and the gap measured 
 in momentum space as a function of the direction of the Fermi momentum
has four nodes along the diagonals  in the Brillouin zone (BZ). The
 nodes have been explicitly detected in 
 angle-resolved  photoemission (ARPES) measurements.~\cite{ARPES_CU} 
 FeSC have multiple FSs  leading to complex 
doping trends and
 rather
 unconventional gap structures for a given symmetry. 
 Still,  ARPES measurements showed~\cite{ARPES_FE} that the
 gap on the FSs centered at $\Gamma$ is
 near-isotropic, clearly inconsistent with $d-$wave symmetry.  

\begin{figure}[tbp]
\includegraphics[angle=0,width=\linewidth]{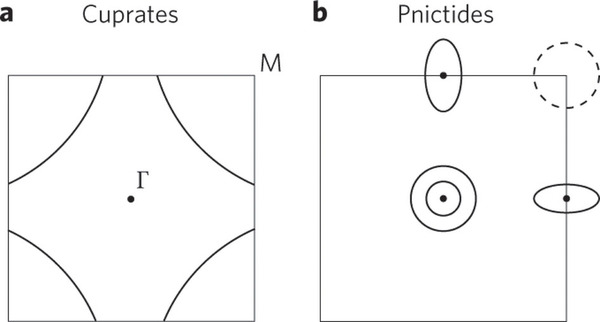}
\caption{Schematics of two-dimensional cros-sections of Fermi surfaces (FSs)
 for cupates and pnictides. 
For weakly doped cuprates, the FS has a single sheet, and filled states
 (the ones closer to $\Gamma$ point) occupy about a half of BZ area. In pnictides, the FS consists of multiple sheets -- 2 hole pockets centered at $\Gamma$, two elliptical electron pockets centered at $(0,\pi)$ and $(\pi,0)$. There is another hole pocket at $()\pi,\pi)$, which is cylindrical in some systems, and near-spherical around $k_z =\pi$ in others~\cite{review_3}.}
\label{fig2}
\end{figure}
Furthermore, the phase diagram of cuprates is richer than just antiferromagnetism and superconductivity -- there is a region of
  Mott insulating behavior at the smallest dopings~\cite{optics_mott} and the 
 still mysterious pseudogap phase occupying a substantial portion of the 
normal state phase diagram, particularly in the underdoped regime 
 (the term  "pseudogap" is commonly used to describe a partial gap of fermionic excitations not accompanied by any obvious long range order).
 A comparison with FeSC again shows differences: there is no Mott phase in undoped pnictides, which show bad metallic, but still metallic behavior of resistivity. As of now, there is 
 only sporadic
evidence of the pseudogap~\cite{optics_fe}. 

The geometry of the FS and the low-energy excitations
 in CuSC and FeSC 
  are also quite different. 
In CuSC, there is a single ``open'' cylindrical FS; its 2D cross-section uncovers four large segments (Fig. \ref{fig2}a). 
 In FeSC, the FS has multiple quasi-2D sheets 
 due to  hybridization between all five $Fe$ d-orbitals --
 there are two small elliptical 
electron pockets centered at $(0,\pi)$ and $(\pi,0)$
  two small near-circular hole pockets 
 centered at $\Gamma$-point (Fig. \ref{fig2}b), and, in some materials, additional hole pocket centered at  $(\pi,\pi)$ 
 The actual FS geometry is even more complex because of extra hybridization 
 due to $Fe-Fe$ interaction via a pnictide/chalkogenide (Ref. \cite{review_3}).
  Given all these disparities in the FS structure, magnetism, and the order parameter symmetry, it is tempting to conclude that the phase diagrams of FeSC and CuSc are merely accidental lookalikes. 
We argue below that the actual situation is more involved. Differences apart,  CuSC and FeSC reveal strikingly similar trends consistent with the notion of all-electronic scenario of electron pairing.  By analyzing these trends we speculate on essential ingredients for unconventional and higher Tc superconductivity.

\section{elements of the pairing mechanism for $Cu$SC and $Fe$SC}

\subsection{gap symmetry and the pairing glue}

A generic pairing scenario for moderately interacting itinerant
 systems assumes
 that fermions attract each other by exchanging quanta of bosonic 
 excitations. A boson can  be a phonon or it can be a collective density-wave
 excitation in either spin or charge channel. In the latter case,
  a  direct
 interaction between the two given fermions is
 purely repulsive, but once it is renormalized by screening and by exchanges with othrer fermions, it acquires complex dependence on the angle along the FS.
 Its overall sign doesn't change, but
  one or more angular momentum components may become attractive. The beauty of superconductivity is that it develops even if 
  just one angular component is attractive, no matter how strongly repulsive are the others. 

In CuSC, there is no consensus on the pairing mechanism, but 
 the most frequently discussed scenario for $d_{x^2-y^2}$ pairing in the optimally doped and overdoped regime is the exchange of collective excitations in the spin channel, commonly referred to
 spin fluctuations~\cite{scal_pines_chub}. 
 Because antiferromagnetic phase is nearby, 
 interaction mediated by spin fluctuations is
 peaked at momenta at or near $(\pi,\pi)$, which links
 fermions in different ``hot regions'' of the BZ near $(0, \pi)$, $(\pi, 0)$, 
  and symmetry related points (Fig. 3a).
  The overall sign of such interaction is positive (repulsive),
 but its d-wave component is attractive, because a d-wave gap changes sign in between hot regions. In FeSC,   ``hot regions'' are replaced by electron  pockets
 centered at  $(0,\pi)$ and $(\pi, 0)$.
  If the pairing interaction were to be  peaked at  $(\pi, \pi)$,
 it would give rise to a  $d-$wave superconductivity with sign-changing gap between electron pockets, in complete analogy with CuSC. This 
    may be the case for recently discovered 
  strongly electron-doped Fe-chalkogenides $AFe_2Se_2$ ($A = K, Rb, Cs$)~\cite{theory_AFeSe}, but for other FeSCs direct interaction between electron pockets is weak, and is overshadowed by the effective interaction 
 through  virtual hoppings to hole FSs~\cite{mazin_kuroki,RPA_Fe,cee,fRG,thomale}. 
 A simple exercise in quantum mechanics shows that such an effective interaction scales as $U^2_{eh}$, where  $U_{eh}$ is electron-hole interaction, and
 is {\it attractive}, i.e., it  gives rise to 
an  $s-$wave pairing (no sign change of the gap between electron pockets),
 in clear distinction  to $d_{x^2-y^2}$ pairing in CuSC.
 
The difference in the gap symmetry does not imply 
  different pairing mechanisms. Indeed, the distance between hole and  electron FSs is the same $(\pi,0)$ or $(0,\pi)$ 
 as the momentum of the stripe magnetic order 
 (this can be easily understood if magnetism is viewd as itinerant~\cite{eremin}).  If $U_{eh}$ is positive (repulsive), stripe magnetic fluctuations do enhance
 $U_{eh}$ and from this perspective provide the glue for s-wave pairing in 
FeSC in the same way as $(\pi,\pi)$ magnetic fluctuations provide for
 the glue for $d-$wave pairing in CuSC. 
  Essentially the same conclusion follows from  RG studies
~\cite{cee,fRG,thomale} which use a somewhat different assumption that the ``glue'' and superconductivity originate from the same 
set of interactions and analyze how spin 
 and pairing susceptibilities {\it simultaneously} grow upon the system's flow towards low energies.  We see therefore that pairing symmetries are
 different in CuSC and FeSC, but the pairing mechanism is 
 likely to be the same -- an exchange of spin fluctuations.

Alternative scenarios for pairing in both CuSC and FeSC postulate dominance of charge fluctuations and/or phonons~\cite{devereaux,kontani}.
 For FeSCs this scenario can be realized~\cite{kontani} provided $U_{eh}<0$ 
(this requires inter-orbital repulsion to be larger than 
intra-orbital one).  For negative  $U_{eh} <0$, stripe magnetic fluctuations are inactive, but charge (orbital) fluctuations and/or phonons
 do  enhance $|U_{eh}|$ and therefore may act as a glue.  We note, however, that   electron-phonon interaction in FeSCs can enhance $U_{eh}$ but is
 too weak to produce a significant $T_c$ 
 on its own~\cite{phonons} 

 Another subtle issue concerns the interplay between 
 pairing fluctuations and fluctuations of the bosonic glue.
  The very idea of a superconductivity mediated by 
a bosonic glue implies that the glue is pre-formed, i.e., 
 bosonic excitations develop at energies well above the ones relevant to superconductivity.  It may be, however, that
 the ``glue'' (e.g., spin fluctuations)
 and superconductivity are caused by the same interactions and develop simultaneously. This last assumption is the base  for weak coupling RG studies~\cite{cee,fRG,zlatko} which analyze how spin and charge susceptibilities and the pairing susceptibility simultaneously evolve upon the system's flow towards 
low energies.  The key physical effect  is the same in both cases --
 pairing susceptibility get enhanced due to coupling to  
 fluctuations of the ``glue'', but the value of the upper cutoff for the pairing (the ``Debye frequency'') is different. When the bosonic glue is preformed, 
the role of a Debye frequency is played by a typical
 frequency of a dynamic spin susceptibility 
(e.g., typical relaxation frequency for spin or charge fluctuations), while 
 if the ``glue'' and superconductivity develop together, the
 ``Debye frequency'' is the scale at which the pairing susceptibility 
 become affected by  the coupling to static fluctuations of a ``glue''.
 In CuSC,  magnetic fluctuations are peaked at or near $(\pi,\pi)$ 
  up to  $200-300 meV$, which is well above the pairing scale, so the ``glue'' may be considered as preformed,  at least for a magnetic scenario.
In FeSC  magnetic fluctuations have been detected up to $200 meV$,  but
 magnetic response at such energies is quite broad in $k-$space, and it still remain to be seen whether high-energy spin fluctuations contribute to a pairing glue or are featurless background which doesn't affect the pairing.   

\subsection{the gap structures}

In CuSC, the geometry of the FS, which consists of a single sheet, and the $d-$wave symmetry  predetermine the momentum dependence  of the superconducting gap$\Delta ({\bf k})$ along the FS -- it changes sign twice and
 has four nodes along the diagonal directions in the BZ. 
 In FeSC, the multiple FSs and 
multi-orbital nature of excitations make the gap structure rather complex,
 even though from the the symmetry perspective it is the simplest  $s-$wave. 

Let us elaborate on the above complexity. First, if the pairing glue are stripe spin fluctuations, an
 s-wave gap adjusts to a repulsive $U_{eh}$ and changes
sign between hole and electron pockets~\cite{review_3}. Such a state is 
 referred to as an extended $s-$wave, or an $s^{\pm}$ state. If
 the pairing is due to orbital fluctuations, the gap is a conventional $s-$wave, or $s^{++}$.
Second, an inter-pocket electron-hole interaction  competes with
 intra-pocket hole-hole and electron-electron repulsions which disfavor any gap with $s-$wave symmetry. Third, both intra-pocket and inter-pocket interactions generally depend on the angles along the FSs.
Because of the last two effects,  $\Delta ({\bf k})$ necessary acquires 
 some angular dependence  to minimize the effect of intra-pocket repulsion and
 to match  angle dependences of the interactions~\cite{RPA_Fe,cee,fRG,thomale}.
  If this angular dependence becomes sufficiently large, the gap develops ``accidental'' nodes at some points along the FSs. Calculations show~\cite{RPA_Fe,cee,fRG,thomale} that the nodes likely develop
 for electron-doped cuprates (the larger the doping, the more probable is 
that there are nodes), while
 for hole-doped FeSC the additional hole FS stabilizes a no-nodal gap. 

Putting subtle issues aside, we see that there are two viable  scenarios for the pairing in FeSC. First is   $s^\pm$ pairing due to attractive interaction between electron pockets and repulsive interaction between hole and electron pockets, and the second  is $s^{++}$ pairing due to attractive interaction between electron pockets {\it and} between electron and hole pockets. These  scenarios yield different gap symmetry compared to that in CuSC, but the pairing mechanisms are essentially equivalent to spin-fluctuation and charge-fluctuation/phonon mechanisms for CuSC. A $d-$-wave pairing in the FeSC is possible for very strongly electron and hole doped
FeSCs, but it
 has been ruled out by the ARPES data for  
FeSCs which contain both hole and electron FSs~\cite{ARPES_FE}. 
 There are some hints~\cite{buechner} for the p-wave gap in in one of FeSC (LiFeAs) but solid evidence is still lacking.

The methods used to determine symmetry and structure of superconducting gaps in FeSC
  were for the most part developed or refined to address the symmetry issue in CuSC. A casual sampling of data acquired with all these techniques --
 neutron resonance~\cite{resonance_Fe}, 
 quasiparticle interference~\cite{hanaguri}, penetration depth~\cite{ pen_depth_Fe}, thermal conductivity~\cite{th_cond_Fe}
  may signal a rather controversial situation on the issue of the gap 
  structure in FeSC unlike the cuprates where the $d_{x^2-y^2}$ state has been ironed out.  Yet, an in-depth query shows that seemingly conflicting results for FeSC are all in accord with the picture of the $s^{+-}$ gap
 by taking proper account of peculiarities of the multiband/multigap nature of this class of compounds.~\cite{RPA_Fe,cee,fRG,thomale,co-existence_Fe_th}.
 Kontani and his collabotators argue, however,
 ~\cite{kontani} that the data do not rule out 
 an $s^{++}$ gap.

 Overall, 
 an importnat lesson learned from the pnictides is that a high symmetry state, {\it e.g., $d_{x^2-y^2}$} is not  a necessity
 to overcome repulsion within an all-electronic mechanism, and that $s$-wave superconductivity is a viable option 
for an electronic pairing in a multi-band high-$T_c$ superconductor.   

\section{Essential ingredients of  high-Tc superconductivity}

We now discuss a number of universal trends detected in Fe-based and
 Cu-based systems. We first point out that  optimally doped 
FeSC and CuSC exemplify conductors in which 
 electrons are neither fully itinerant nor  completely localized~\cite{qazilbash09,si} . 
 A way to quantify the tendency towards localization is to analyze the 
 experimental kinetic energy $K_{exp}$ that can be determined from ARPES or
  optical measurements in conjunction with the non-interacting value 
 provided by band structure calculations, $K_{LDA}$~\cite{millis05,qazilbash09,Haule-Kotliar-nat-phys,shen-lafepo}. 
 The two extremes of the $K_{exp}/K_{LDA}$ ratios are instructive. 
 In conventional metals and electron-phonon superconductors $K_{exp}/K_{LDA}\simeq 1$ signaling that there is not need to invoke  
localizing trends to explain electrons dynamics.
 In the opposite extreme of Mott insulators,
 localization arrest electron motion and  $K_{exp}/K_{LDA}\rightarrow 0$. 
To the best of our knowledge, in {\it all} exotic superconductors,  including both FeSC and CuSC, there is a noticeable renormalization of the kinetic energy, i.e., all these systems show some tendency towards localization
 (see Table \ref{Table1}).
Notably,  for materials with the highest $T_c$ in each family,  $K_{exp}/K_{LDA} \sim 0.3-0.5$ implying a substantial distance away from 
 both  purely itinerant and Mott regimes. We therefore witness a remarkable tendency of 
materials with the highest $T_c$ for a given series  to strike the right balance between considerable strength of 
 interactions and itinerancy. 
Because an estimate of $K_{exp}$ is readily accessible from experiments at ambient conditions\cite{qazilbash09}, the above rather remarkable unifying aspect of a extremely diverse group of superconductors can be exploited for the search for new superconducting materials. 

Why  must  $K_{exp}/K_{LDA}$ be ``right in the middle''
 to yield a high $T_c$? We believe the most generic reasoning for this
 ``Goldie Locks law of superconductivity'', is rather straightforward.
  At weak coupling, itinerant fermions  are ready to superconduct upon pairing, but $T_c$ is exponentially small. In the opposite limit, when the 
 interaction strength exceeds the fermionic bandwidth,
 fermions are completely localized and cannot move, even 
though the binding gap $\Delta$ in this latter case is large. 

A connection between the itinerancy-localization balance and the superconducting $T_c$ can be appreciated by realizing that 
 our measure of the interaction strength
 $K_{exp}$ also sets the upper limit for the superfluid density $\rho_s$.
  Once $K_{exp}$ is diminished, so is $\rho_s$. Without proper superfluid stiffness superconductivity becomes susceptible to the destructive role of phase and amplitude gap fluctuations~\cite{emery-kivelson-prl95} and of competing orders. As a consequence, $T_c$ is reduced compared to 
  $\Delta$. A necessity for substantial $\rho_s$ (and therefore not too small
 $K_{exp}$) is epitomized through "the Uemura plot": $T_c\propto \rho_s$,
 which holds for all exotic superconductors.\cite{Uemura-prb10} 

\begin{table}[htp]
\centering
\begin{tabular}{|c|c|c|c|}
\hline
Superconductor&$T_{c,max}$&$K_{exp}/K_{LDA}$&Refs\\
\hline
{\bf CuSCs}&&&\\
\hline
 $Nd_{2-x}Ce_xCuO_4$&35&0.3&\cite{millis05} \\
\hline
 $Pr_{2-x}Ce_xCuO_4$&35&0.32&\cite{millis05} \\
\hline
 $La_{2-x}Sr_xCuO_4$&40&0.25&\cite{millis05} \\
\hline
 $YBa_2Cu_3O_7$&93.5&0.4&\cite{optics_mott} \\
\hline
 $Bi_{2}Sr_2CaCu_2O_8$&94&0.45& \cite{vdm-private}\\
\hline
\hline
{\bf FeSCs}&&&\\
\hline
 $LaFePO$&7&0.5&\cite{qazilbash09} \\
\hline
 $Ba(Fe_{1-x}Co_x)_2As_2$&23&0.35-0.5&\cite{qazilbash09,schafgans11} \\
\hline
 $Ba_{1-x}Co_xFe_2As_2)$&39&0.3&\cite{charnukha} \\
\hline
\hline
{\bf Exotic SCs}&&&\\
\hline
 $CeCoIn_5$&2.3&0.17& \cite{singley,shim}\\
\hline
 $Sr_2RuO_4$&1.5&0.4&\cite{qazilbash09} \\
\hline
 $\kappa-(BEDT-TTF)_2Cu(SCN)_2$&12&0.4&\cite{basov_rmp} \\
\hline
\hline
{\bf Electron-phonon SCs}&&&\\
\hline
 $MgB_2$&40&0.9& \cite{qazilbash09}\\
\hline
 $K_3C_{60}$&20&0.96&\cite{degiorgiC60,erwin} \\
\hline
 $Rb_3C_{60}$&30&0.9&\cite{degiorgiC60,erwin} \\
\hline
\end{tabular}
\caption{
The ratio of the  experimental kinetic energy $K_{exp}$ extracted
 from ARPES and optical measurements, as described in Ref.\protect\cite{qazilbash09},  and $K_{LDA}$ provided by band structure calculations.}
\label{Table1}
\end{table}

 \begin{figure}[tbp]
\includegraphics[angle=0,width=\linewidth]{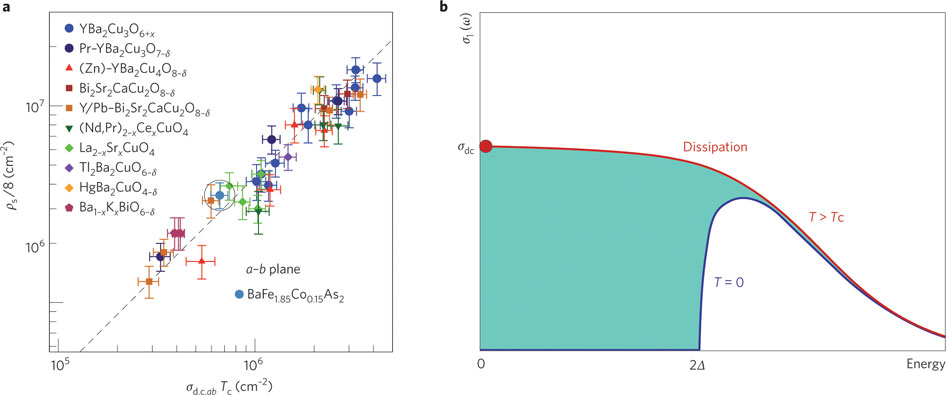}
\caption{Superfluid density in exotic supercondutors. Left panel: Homes 
 scaling of the superfluid density $\rho_s$ versus the product of the DC conductivity and the transition temperature $\sigma_{DC}\times T_c$ 
 for several families of exotic superconductors including cuprates and pnictides (from Ref.\protect\cite{tu}). Right panel:
schematics of the condensate formation in a superconductor. The superfluid density is given by the “missing area” in the spectra of the real part of the conductivity $\sigma_1(\omega)$ primarily from the region given by the energy gap $2\Delta$. This area can be estimated from the product of 
$\sigma_{DC} \times 2\Delta$.}
\label{fig4}
\end{figure}
Is  $K_{exp}/K_{LDA}$ the only parameter 
 essential for $T_c$? No. We argue that $T_c$ can be further modified  even when
 $K_{exp}/K_{LDA}$ is fixed at ``optimal'' value by changing the structure of low-energy fermionic excitations inside the band. 
 An important input for this consideration is the empirically determined ``Homes scaling''  $\rho_s\propto \sigma_{DC}\times T_c$\cite{homes2004,Homes-prb10,tu} where $\sigma_{DC}$ is the DC conductivity just above $T_c$. This scaling law holds for both  FeSC and CuSC (see Fig. \ref{fig4})
 and establishes a
  link between superconductivity and the normal state transport.
 Sum rules provide useful guidance to appreciate this link.
  According to the Ferrell-Glover-Tinkham sum rule,
 the superfluid density is given by the missing spectral weight in the real 
part of the conductivity.
 In a superconductor with  strong dissipation 
 the conductivity spectra in the normal state are broadened by scattering.
 In panel b of Fig. \ref{fig4} we 
 show schematically 
 that in this case the magnitude of $\rho_s$ can be reliably estimated by taking a product of $2\Delta\times \sigma_{DC}$. 
Because $\Delta \propto T_c$, the Homes scaling holds.
In BCS-type superconductors the key source for dissipation is disorder.
 A common and highly non-trivial aspect of both FeSC and CuSC in this context
 is that strong dissipation has little to do with disorder, as evidenced by observations of quantum oscillations  which demand high purity of the  specimens~\cite{amalia}. Instead, strong dissipation in both FeSC and CuSC is an inherent characteristic of charge dynamics at finite frequencies. 
In many cases, the dissipation is caused by the same interaction (e.g., spin-fluctuation exchange) that gives rise to the pairing.
Infrared and ARPES measurements support the notion of strong dissipation by registering incoherent spectral features\cite{ARPES_CU,Dressel-prb09,optics_fe,vdm2003}. A transport counterpart of these effects is the linear 
temperature dependence of the resistivity above $T_c$ 
 and  $T^{\alpha}$ behavior of resistivity, with $1<\alpha <2$, down to $T=0$ at the end point of superconductivity in the overdoped regime~\cite{taillefer}

 Strong dissipation suppresses  fermionic coherence and, at a first glance, should also diminish the ability of fermions to (super)conduct. 
 However, strong dissipation does not necessary require the interaction to be larger than the bandwidth as it can be additionally enhanced by bringing the system to the vicinity of a quantum-critical point. In this latter case,  
 the fermionic self-energy 
$\Sigma (\omega,k)$ at energies below (already renormalized) 
electronic bandwidth 
becomes predominantly frequency-dependent, what
makes fermions incoherent but does not localize them.
 Furthermore, $T_c$  actually {\it increases}  
 with increasing   $\Sigma (\omega)$ because the suppression of fermionic coherence is overshadowed by the simultaneous increase of the dynamical pairing interaction. 
 The increase is stronger in quasi 2D systems than in 3D.~\cite{lonzarich}
In CuSC, the effect of dissipation has been analyzed from various perspectives and the upper limit on $T_c$ was found to be around 2\% of $E_F \sim 1 eV$,
 (Ref.\cite{abanov}), in good agreement with the experimental $T_c$ values.  
 Full scale calculations of $T_c$ in FeSC have not been done yet and 
are clearly called for.

We summarize this article by a brief outline of common characteristics of
pnictides and cuprates that may teach us a lesson on generic attributes of
a high-$T_c$ superconductor and thus may aid the search for new materials
with even higher $T_c$. First, the 
screened Coulomb interaction 
should be strong, but not too strong to 
induce localization causing the reduction 
of $\rho_s$. 
The interaction of the order bandwidth appears to be optimal leading to $K_{exp}/K_{LDA}\simeq 0.5$. Second, the compliance of exotic and high-Tc superconductors with the Homes law demands  strong
 dissipation
that can be registered through transport and spectroscopic methods.
This dissipation predominantly comes from the effective dynamical electron-electron interaction within the band rather than from disorder. The same dynamical interaction
gives rise to the pairing.
  Third,
it is imperative that a system is able to avoid the repulsive nature of
 electron-electron interactions.
 Cuprates and pnitides have instructed us that
there is more than one way to deal with the repulsion problem ($d-$wave
pairing in the cuprates and gap variations between multiple FSs in the
pnictides). Finally, we stress significance of real space inhomogeneities 
that may in fact favor the increase of $T_c$ under optimal
conditions~\cite{kivelson}. Notably, ALL these effects are observed both
in Fe-based and Cu-based systems thus identifying to a surprisingly
consistent leitmotif of high $T_c$ superconductivity driven through
all-electronic interaction in these systems. A theoretical challenge is to
accommodate these diverse effects in a microscopic theory with a
predictive power. On a practical side, incorporating the above
prerequisites into a viable protocol that facilitates the search for new
superconductors is still an iron in the fire.

\section*{Acknowledgements}
The authors wish to thank for valuable discussions large numbers of colleagues 
 working on both cuprates and pnictides.  This work was supported by 
 NSF, DOE and AFOSR (D.N.B) and by 
NSF-DMR-0906953 (A.V.C).

\end{document}